# A Method for Target Detection Based on Mmw Radar and Vision Fusion

**Ming Zong, and Jiaying Wu, Student Member, IEEE, Zhanyu Zhu, and Jingen Ni, Senior Member, IEEE**

Abstract—An efficient and accurate traffic monitoring system often takes advantages of multi-sensor detection to ensure the safety of urban traffic, promoting the accuracy and robustness of target detection and tracking. A method for target detection using Radar-Vision Fusion Path Aggregation Fully Convolutional One-Stage Network (RV-PAFCOS) is proposed in this paper, which is extended from Fully Convolutional One-Stage Network (FCOS) by introducing the modules of radar image processing branches, radar-vision fusion and path aggregation. The radar image processing branch mainly focuses on the image modeling based on the spatiotemporal calibration of millimeter-wave (mmw) radar and cameras, taking the conversion of radar point clouds to radar images. The fusion module extracts features of radar and optical images based on the principle of spatial attention stitching criterion. The path aggregation module enhances the reuse of feature layers, combining the positional information of shallow feature maps with deep semantic information, to obtain better detection performance for both large and small targets. Through the experimental analysis, the method proposed in this paper can effectively fuse the mmw radar and vision perceptions, showing good performance in traffic target detection.

## I. INTRODUCTION

Traffic intersection is one of the most challenging surveillance scenarios in smart transportation of urban environment, requiring reliable and accurate sensing information to ensure the traffic safety. However, vision camera using the ambient illumination light is the major sensor in traffic intersections [1]. However, in special situations such as rainy, snowy, and foggy weather, the effectiveness of visual sensors will be greatly limited, which requires the fusion of multiple sensors to improve the accuracy of target detection. Millimeter-wave (mmw) radar has the ability to provide accurate range and velocity measurement all weather conditions, while visual sensors just acquire the vision information depends on the light condition . When fusing these two sensors, it is possible to compensate for their shortcomings respectively and improve the monitoring accuracy of traffic intersections [2].

Mmw radar is one of the most fundamental and important sensors in roadside sensing systems, which uses the Doppler effect to determine the relative parameters of the detected target (such as distance, velocity, and angle) [3]. The operating frequency range of mmw radar is from 30GHz to 300GHz, which is widely used in the practical applications. In literatures [4] and [5], a mmw radar sensor using frequency modulated continuous wave (FMCW) is introduced, which uses two Fast Fourier Transforms (FFTs) to obtain the temporal and spatial dimension information of the signal, achieving all-weather detection and tracking of targets in the field of view. This is also the principle basis of mmw radar target detection. In order to reduce error and redundancy checks, additional filtering and clustering methods are generally used to correct radar measurement data [6]. Classical density-based method to cluster real targets and remove false targets is also researched [7]. In recent years, deep neural networks have begun to play an important role in the field of radar detection [8], including Convolutional Neural Networks (CNNs) [9], Recurrent Neural Networks (RNNs) [10], and Long Short-Term Memory (LSTM). Many neural network technologies based on the Lidar detections, such as PointNet and PointNet++, have extended corresponding versions for mmw radar. However, since the sparsity of mmw radar point cloud and the influences of random noise , it is difficult to detect traffic targets accurately only use mmw radar in roadside surveillance, especially in terms of shape perception and other detail information [11].

As one of the most common sensors at traffic intersections, cameras have advantages on low cost and strong shape perception ability. They can capture high-resolution images, videos, etc., and have outstanding advantages on target recognition. With the development of image processing technology based on deep learning, vehicle recognition detection based on vision has become the research hot spot [12]. The target detection based on CNNs can be mainly divided into two categories: two-stage detectors and one-stage detectors. Two-stage detection refers to the two steps of the network: candidate region generation and target classification, which is mainly on behalf of Region-CNN (R-CNN). R-CNN algorithm combines candidate region extraction with CNN, utilizing the feature extraction ability of CNN to improve target detection accuracy. Reference [13] proposed Spatial Pyramid Pooling Networks (SPP-Net), which utilizes SPP to reuse feature maps for candidate area extraction. Similarly, the Fast R-CNN proposed in reference [14] utilizes pooling layers to extract regions of interest into feature maps of any size, thereby achieving feature map reuse. The Fast R-CNN proposed in Reference [15], which adds a Region Proposal Networks (RPN) module composed of convolutional layers on the basis of [14], accelerates the operation speed of the network. After that, more two-stage networks emerges, such as Feature Pyramid Network (FPN), Region-Based Fully Convolutional Network (R-FCN) [16], Mask R-CNN etc. However, because of their large number of parameters, high computational complexity, and slow operating rate, it is difficult to satisfy the processing speed requirements of roadside equipment. Compare with two-stage networks, one-stage networks omit the steps of candidate region generation and upgrade the processing speed of target detection. One-stage networks are represented by the YOLO series [17] and Single Shot Detector (SSD) [18]. The YOLO algorithm utilizes the regression process to directly predict bounds and the degree of confidence on image grid cells, while the SSD algorithm introduces an anchor box mechanism on this basis, cancels the fully connected layer and use multiple feature maps for target detection to achieve a balance between detection speed and accuracy. The FCOS proposed in Reference [19] uses anchor free box to predict pixel points. Reference [19] also proposes a center-ness branch to compensate for the error between the predicted pixel points and the corresponding bounds center. With the development of target detection technology based on vision, the methods above have improved detection accuracy and speed. However, in practical engineering, using a single visual sensor is easy to be interfered from the external environment, such as weather, environment, lighting, etc. Therefore, multi-sensor fusion has become a popular research direction in roadside perception devices.

Currently, there are shortcomings in the research on vehicle detection at traffic intersections in roadside perception. Existing single visual sensors (used in video surveillance) or mmw radars (used in speed measurement radars) cannot maintain high accuracy and robustness in vehicle detection at intersections. Multi-sensor fusion schemes are mostly used in autonomous driving scenes, and some fusion schemes for roadside perception are mainly focused on simple backgrounds such as highways. The multi-sensor perception scheme for complex environments such as urban traffic intersections still needs to be supplemented. In the current fusion algorithm of mmw radars and visual sensors, it is unable to meet the detection of both far and close range targets in bad weather conditions. It is difficult to satisfy the real-time and accurate perception task of roadside perception devices for road traffic targets.

Then a method for target detection based on mmw radar and vision fusion is studied in this paper. The major contributions are summarized as follows: 1) we design a target detection algorithm based on mmw radars and visual sensors; 2) we design a method for target detection using RV-PAFCOS on the basis of FCOS; and 3) we conduct the spatiotemporal calibration of multiple sensors and RGB image modeling of mmw radar point cloud results.

The rest of this paper is organized as follows: Section 2 briefly introduces three modes of multi-sensor data fusion and target detection algorithms based on CNN. In Section 3, based on the target detection algorithm FCOS, a radar and vision fusion module is added to propose the RV-PAFCOS network and describe the design principles of the fusion module in detail. Section 4 describes the spatiotemporal calibration of multiple sensors and RGB image modeling of mmw radar point cloud results in detail, and introduces the dataset used for network training. Section 5 demonstrates the performance of RV-PAFCOS in target detection experiments and evaluate and analyze its detection effectiveness. Section 6 summarizes the work of the full paper.

## II. RELATED TECHNOLOGIES

The target detection algorithm based on mmw radar and visual fusion mainly involves the principle of multi-sensor data fusion and target detection algorithms based on CNN, which will be briefly introduced in this section.

### A. Principle of multi-sensor data fusion

When using multiple sensors for target detection, it is necessary to fuse the information collected by the sensors to obtain more accurate and robust detection results. Currently, sensor fusion is mainly divided into three modes: data level fusion, decision level fusion and feature level fusion. The flowchart is shown in Fig. 1.

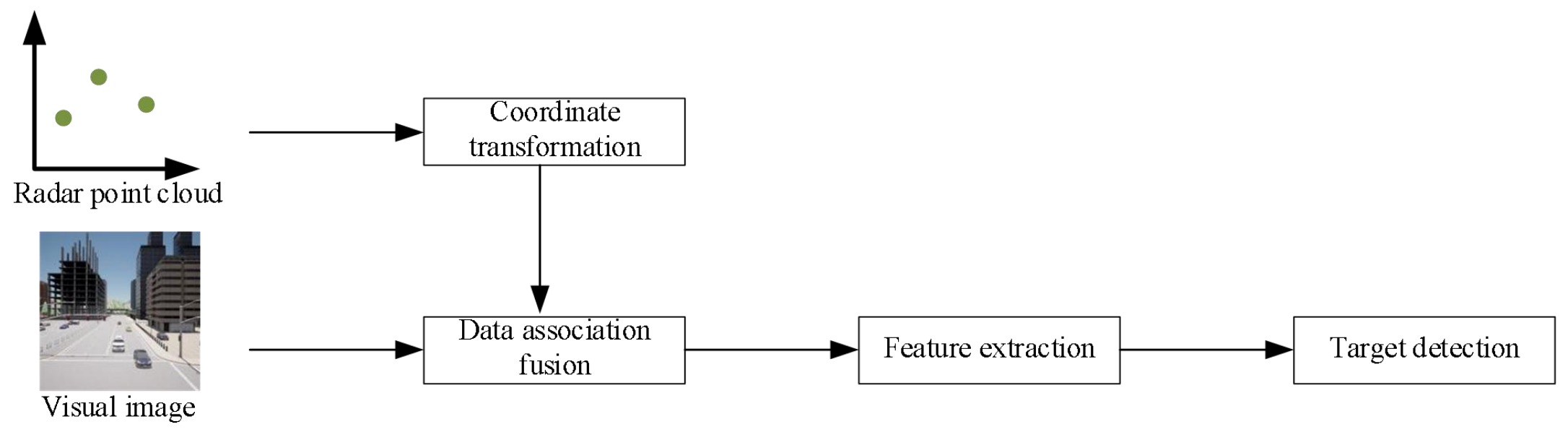

(a)

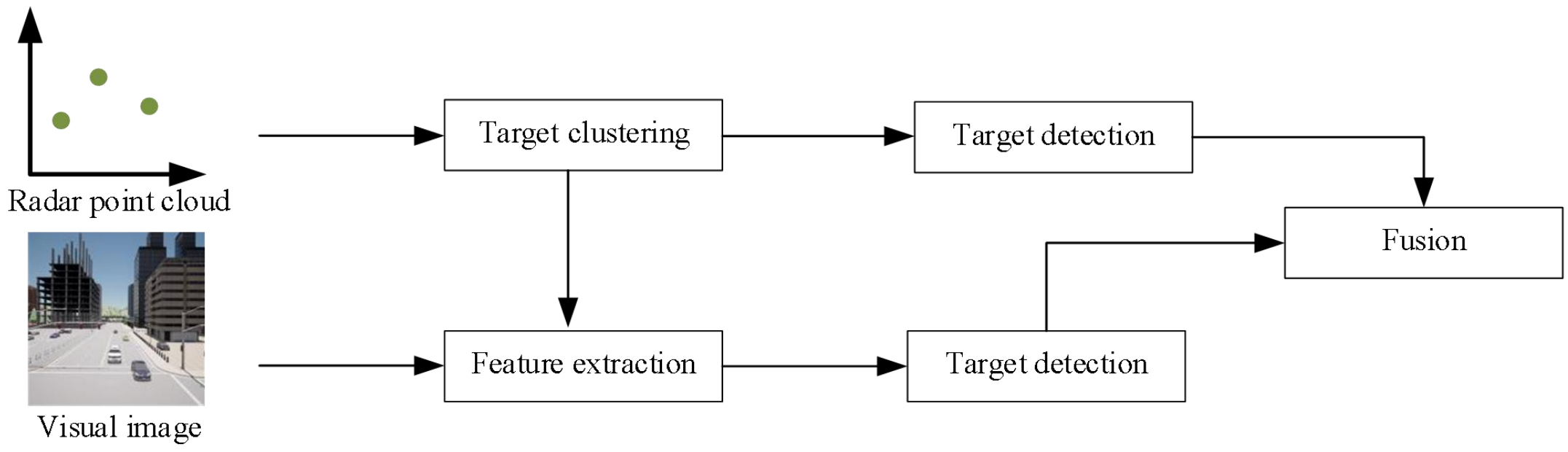

(b)

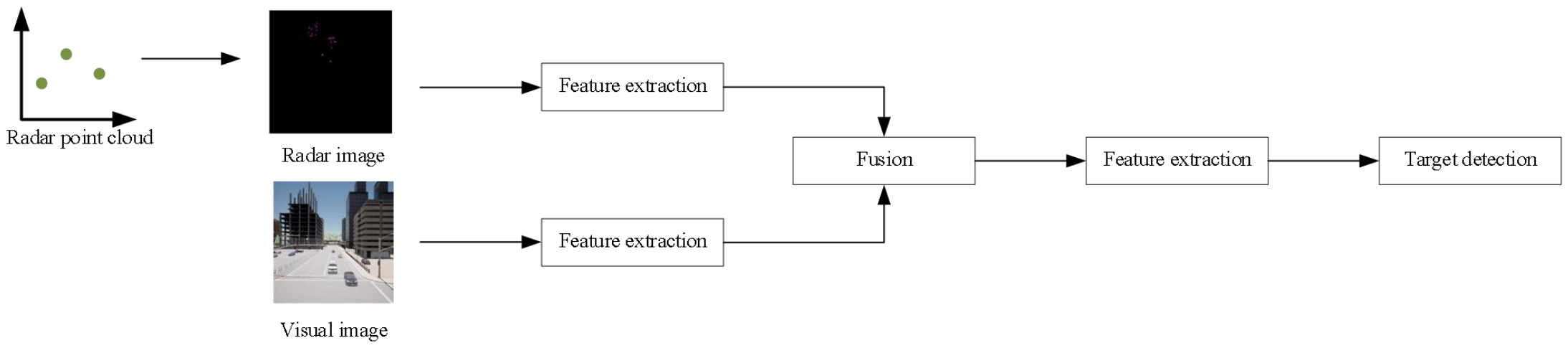

(c)

Fig. 1. Three types of multi-sensor fusion modes: (a) Data level fusion, (b) Decision level fusion, (c) Feature level fusion.

The advantage of feature level fusion is that it can use the characteristics of different sensors to extract more comprehensive information, thus improving the accuracy of detection and classification. It should be noted that feature level fusion should ensure that the feature maps of each sensor have the same size and semantic information, otherwise it will affect the fusion effect. In addition, feature level fusion requires the fusion of multiple layers of features, which also increases computation and complexity.

Table I shows the performance comparison of three fusion modes, among which the detection accuracy of data level fusion is relatively high, but it requires high original data format and is difficult to fuse. Decision level fusion does not require raw data and directly fuses the prediction results of various sensor branches, but the detection accuracy is also relatively low. Feature level fusion achieves a balance between detection accuracy and fusion difficulty. Therefore, this paper fuses mmw radar and visual sensors at the feature level, and the specific work will be introduced in the following paper.

TABLE I

PERFORMANCE COMPARISON OF THREE FUSION MODES

| Fusion modes | Accuracy | Raw data format requirements | Integration difficulty | Computation |
|---|---|---|---|---|
| Data level fusion | High | High | Difficult | Large |
| Decision level | Low | None | Easy | Small |
| feature level fusion | Middle | Low | Middle | Middle |

B. Target detection algorithm based on CNN

CNN, as a powerful image recognition model, is widely used in the field of target detection. Target detection networks can detect and locate multiple targets in an image simultaneously, mainly including feature extraction networks and detection heads. Common target detection networks include Faster R-CNN, YOLO, SSD, etc. All of them use anchor mechanisms and require anchor boxes of different scales on different feature maps in order to achieve good detection results on targets of different sizes. However, this method has some drawbacks, such as increasing training difficulty and computational complexity with too many anchor boxes, and requiring a large amount of hyperparameter adjustment, which is not easy to operate.

In order to improve the efficiency of target detection, the common method is to combine convolutional networks with specific target detection algorithms. Among them, FPN [20] and FCOS [19] are currently popular methods.

FPN is a target detection network composed of feature pyramid networks and feature aggregation networks. The basic idea is to extract multi-scale features of the image through a top-to-bottom path, and then combine shallow features with deep features through horizontal connections to form a cross-scale feature pyramid. This feature pyramid can provide precise localization and abundant semantic information for targets of different scales, thereby improving the accuracy and efficiency of target detection.

FCOS is a target detection method based on anchor-free, which uses an approach based on pixel and does not require predefined anchor boxes. The detection head of FCOS directly classifies and regresses each pixel, and predicts the center point and bounds of the target on the feature map. As a result, it can achieve a single-stage detection process and reduce computational complexity and memory consumption greatly. This method can avoid the problem of requiring a large number of anchor boxes in the anchor-based method and losing information about small targets.

In the field of target detection, FCOS has obtained widespread attention and application because of its excellent single-stage performance and the advantage of not requiring predefined anchor boxes. Compared with traditional anchor box mechanisms, FCOS can adapt to various target scales better. At the same time, it can reduce many hyperparameter adjustments, which makes the entire detection process more efficient. Therefore, this paper researches on target detection algorithms based on FCOS that can achieve multi-sensor data fusion.

## III. TARGET DETECTION ALGORITHM BASED ON RADAR-VISION FUSION

This section proposes RV-PAFCOS for target detection based on mmw radar and visual fusion. This network adds the Radar-Vision Fusion module on the basis of FCOS to integrate image features captured by millimeter wave radar and visual sensors. Distant vehicles have fewer pixels in the image and belong to small targets, which can only be recognized through shallow feature maps with higher resolution. On the other hand, nearby vehicles have more pixels in the image and belong to large targets, which can be recognized through deep feature maps. Therefore, the idea of PANet [21] is introduced into the network, which considers feature layers for bottom-to-top path enhancement and combines shallow positional information with deep semantic information to achieve good detection performance for both large and small targets. The overall detection framework of RV-PAFCOS is shown in Fig. 2.

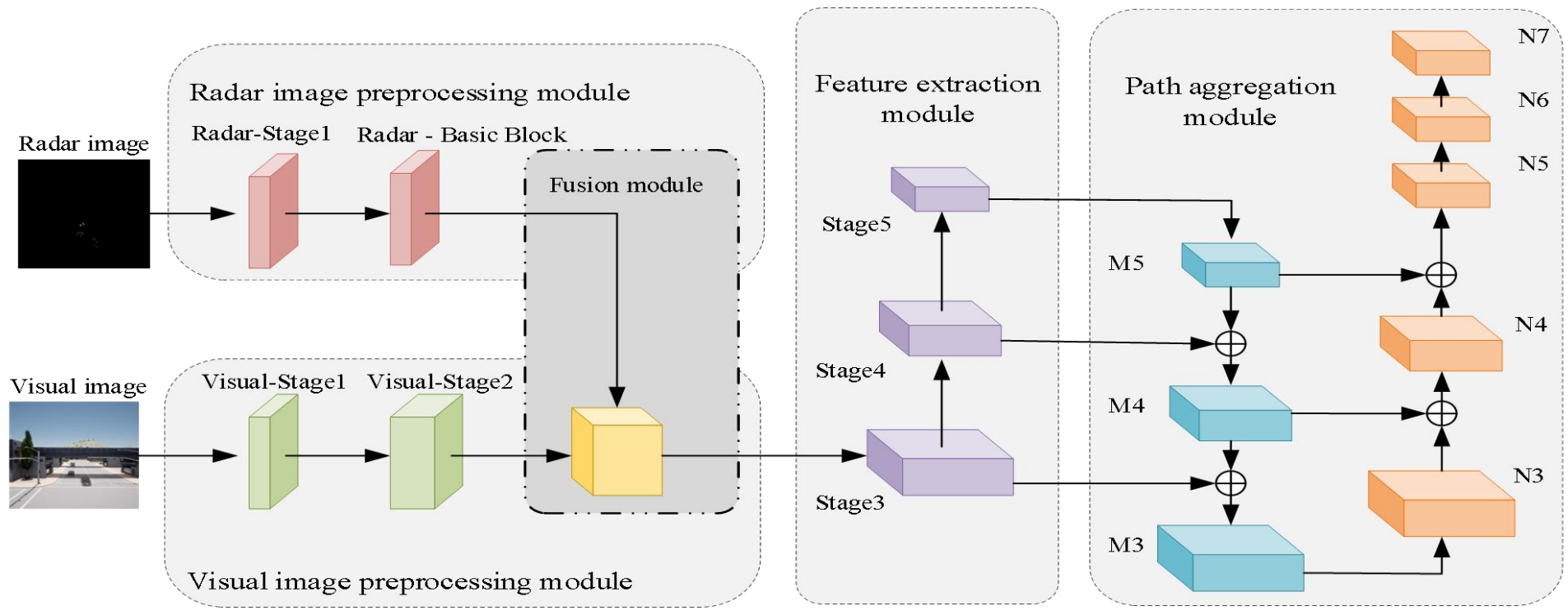

Fig. 2. Network architecture of RV-PAFCOS.

## A. Network detection framework

This section proposes a feature fusion detection network based on the FCOS framework, which mainly consists of five parts: radar image preprocessing module, visual image preprocessing module, fusion module, feature extraction module and path aggregation module. The radar image preprocessing module and visual image preprocessing module are improved on the basis of ResNet [22]. Taking the radar as an example, they include two convolutional blocks: radar-Stage 1 and radar-basic blocks corresponding to the Stem block and Stage1 of ResNet respectively. The internal convolutional layer network structure is shown in Fig. 3. The meaning of Conv1 3, 64, 7×7, 2, 3 in the picture is a convolutional layer with an input channel of 3, an output channel of 64, a convolution kernel size of 7×7, a stride of 2 and a padding of 3. After two images are input, they both go through the same convolutional layer to process the input data. However, in the second stage, there are differences. The visual branch has three complete residual blocks, while the radar branch only uses the residual blocks from the first layer. The reason is that radar images have sparsity. When there are too many residual blocks set, it is difficult for the detection model to update through random gradient descent. In addition, fewer residual blocks can also save computational resources. The fusion module fuses the feature images extracted from two image branches, mainly including addition, multiplication, concatenation and spatial attention concatenation fusion, etc. Subsequently, through the three residual module groups of the feature extraction module, the fused feature maps of radar and visual sensors are extracted to obtain multi-scale feature maps. The network structure of this part is the same as the residual module group of the ResNet backbone network at the same stage. In the final path aggregation module, the multi-scale feature maps obtained earlier are firstly transmitted with deep features in a top-to-bottom pattern and then enhanced with bottom-to-top paths, which utilize accurate shallow information to enhance the entire feature layers. The specific network structure is shown in Fig. 4.

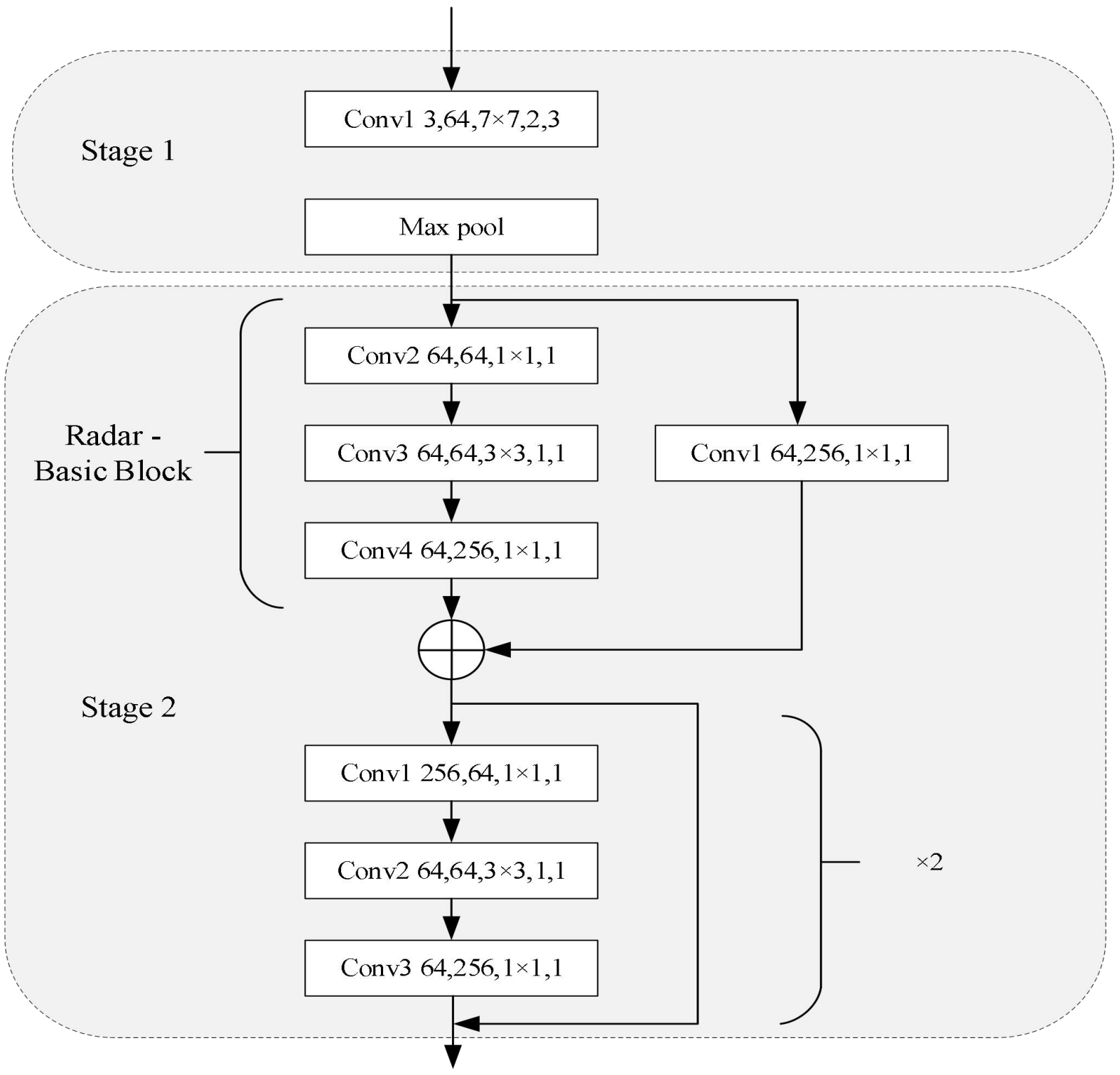

Fig. 3. Convolutional layer structure diagram of the preprocessing module

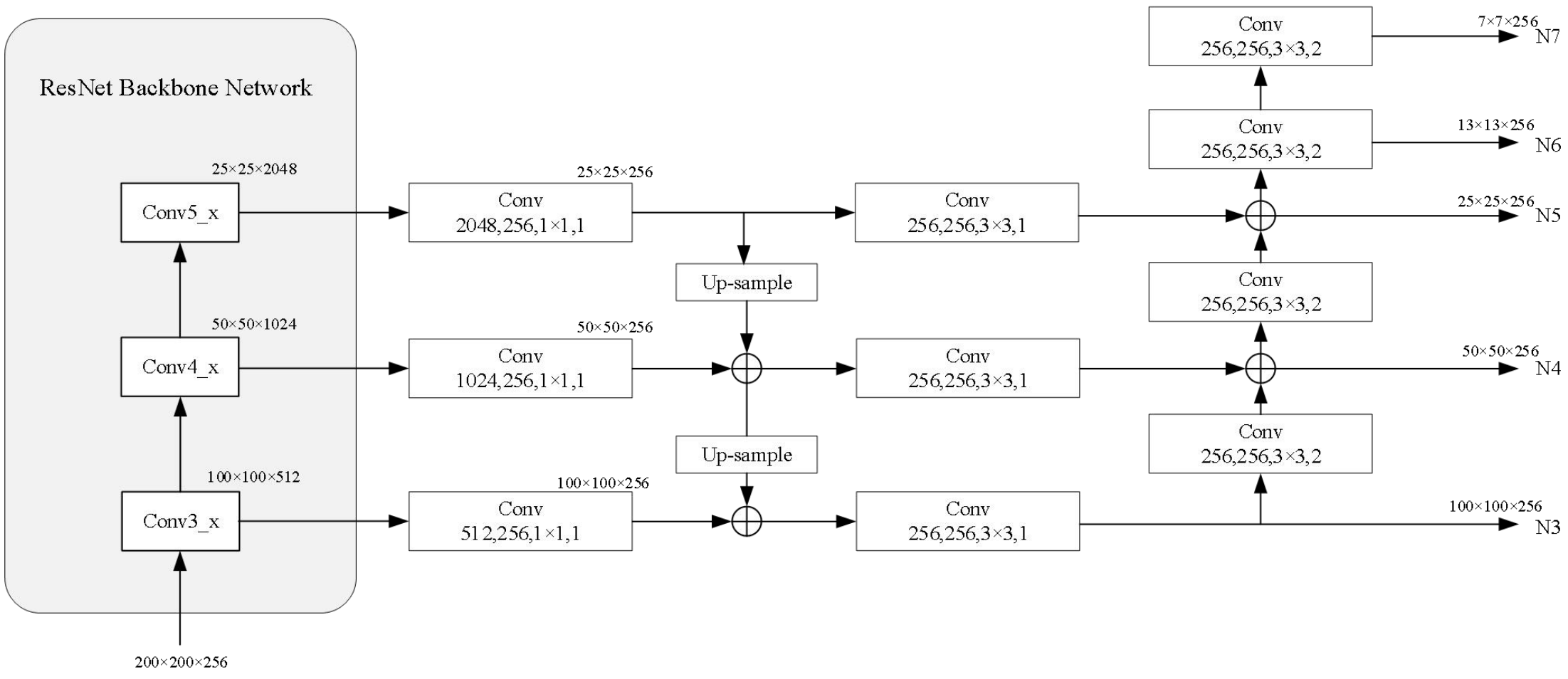

Fig. 4. Network structure of path aggregation module

In the head of the detection section, there are five feature map inputs, namely N3 - N7, where N3, N4 and N5 are feature fusion outputs after path aggregation, while N6 and N7 are feature maps completed by two down-sampling operations and obtained by performing a 2-step convolution on the basis of N5 to obtain deeper feature information. This paper uses the same loss function as FCOS in the detection section [19], consisting of category loss and bounding box regression loss. The formula is shown in (1):

$$L(\{P_{x,y}\},\{t_{x,y}\}) = \frac{1}{N_{pos}}\sum_{x,y} L_{cls}(P_{x,y}, c^*_{x,y}) + \frac{\lambda}{N_{pos}}\sum_{x,y} l_{c^*_{x,y}>0} L_{reg}(t_{x,y}, t^*_{x,y}) \quad (1)$$

where $P_{x,y}$ represents predicted class labels, $c^*_{x,y}$ represents the true value of labels at specific positions (x, y) in the feature map, $t_{x,y}$ represents predicted bounds, $t^*_{x,y}$ represents the true value of bounds, $L_{cls}$ represents Focal Loss, $L_{reg}$ represents GIoU loss, $N_{pos}$ represents positive sample size, $\lambda$ represents the weight of regression loss, whose default value is 1, and $l_{c^*_{x,y}>0}$ represents indicator function. When $c^*_{x,y} > 0$, the function is 1, otherwise it is 0.

B. Radar-Vision image fusion module

In roadside perception systems, mmw radar and visual images are two important perception methods, and they have complementary characteristics. Mmw radar reflects the physical state of targets within the detection range. Using radar points as grids can make the information flow extracted by visual sensors more effective, enhance the detection performance of small and fuzzy objects, and improve the recall rate in detection.

After the mmw radar and visual image preprocessing module in RV-PAFCOS, feature level data fusion is completed by operating on the feature layer. Fig. 5 shows four specific fusion modes, namely pixel level Add Fusion (ADD), Multiplicative Fusion (MUL), Concatenate Fusion (CAT), and Spatial Attention Concatenate Fusion (SAC). Among them, ADD and MUL are the simplest fusion methods, which only require the same size and number of channels of the two types of images. Then, each pixel of the two can be added or multiplied to obtain the fused result. ADD was evaluated in reference [23], and experiments show that it improves detection performance compared with before fusion. In CAT, the RV-Net and CRF-Net proposed in reference [24] uses splicing fusion to connect mmw radar images and visual images by channel in order to form new images.

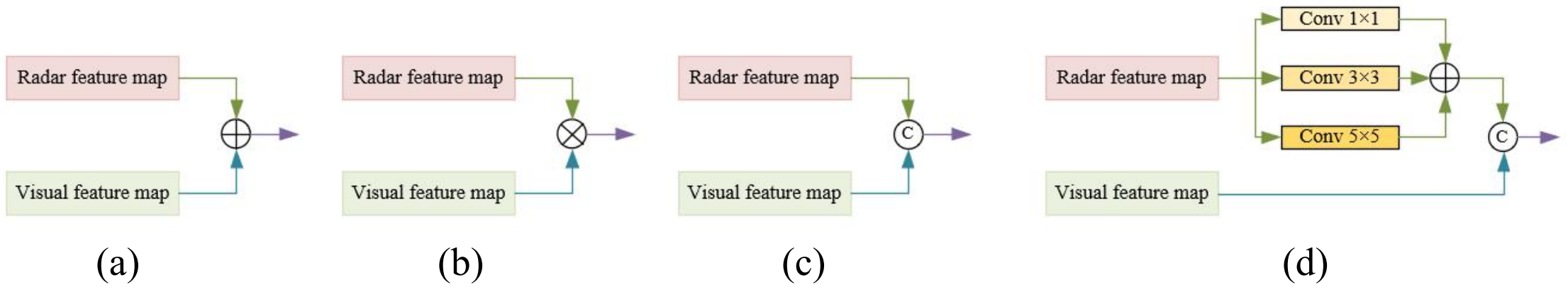

Fig. 5. Four mmw radar and visual image fusion methods: (a) Addition, (b) Multiplication, (c) Concatenate, (d) Spatial attention concatenate.

Spatial attention stitching fusion refers to the fusion of radar images with visual images after extracting spatial attention weight matrices through convolutional layers of different scales. The spatial attention fusion method in this paper mainly considers concatenating the weight matrix with visual features. Taking Fig. 5(d) as an example, Conv 1×1 means going through a convolutional layer with an input channel of 256, an output channel of 1, a convolution kernel size of 1×1, a stride of 1 and a padding of 0, abbreviated as Conv 256, 1, 1×1, 1, 0. Conv 3×3 and Conv 5×5 means Conv 256, 1, 3×3, 1, 1 and Conv 256, 1, 5×5, 1, 2 respectively. The use of convolutional layers with three different kernel scales is to generate a radar attention weight matrix with multi-scale receptive fields, thereby learning the semantic and positional information contained in radar points, and using it as a spatial attention matrix to enhance the corresponding feature information of visual images. Fig. 6 shows the visualization results of mmw radar and visual image feature maps before and after the four fusion methods mentioned above. The first and second columns are the feature maps obtained by the preprocessing module of mmw radar and visual images, and the last four columns are the feature maps obtained by the four fusion methods. For these four methods, experiments will be conducted in the following sections to analyze the accuracy and robustness of the target detection network based on data fusion.

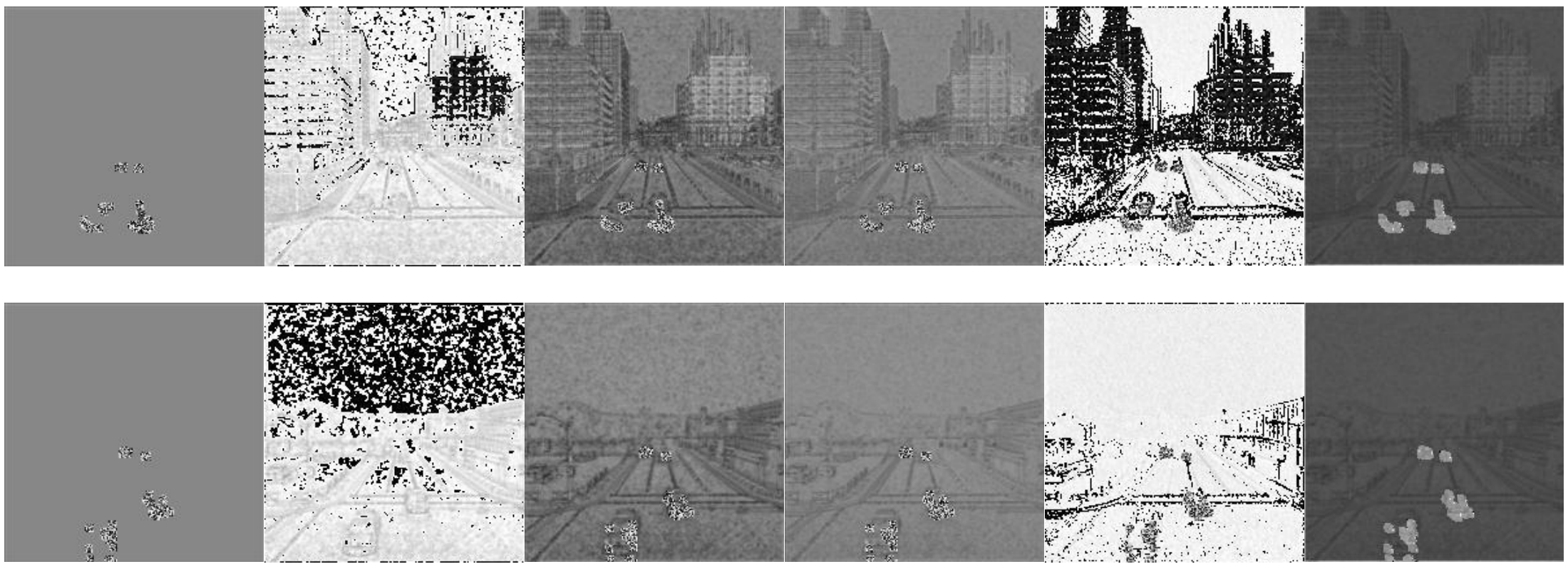

(a) (b) (c) (d) (e) (f)

Fig. 6. Visualization effect of four fusion module feature maps: (a) Radar image, (b) Visual image, (c) ADD, (d) MUL, (e) CAT, (f) SAC.

## IV. DATA PROCESSING BASED ON RADAR-VISION FUSION

The previous section proposed the target detection algorithm RV-PAFCOS for multi-sensor fusion, which requires the radar point clouds to be converted to radar images before radar-vision fusion module, that is, the target point cloud data is projected onto the same image plane as the visual image. So, this section firstly performs spatiotemporal calibration on the two sensors to achieve spatiotemporal alignment between them, then constructs a mathematical model for converting mmw radar point cloud data into RGB images, and finally introduces the dataset for training and analysis on RV-PAFCOS.

### A. Multi-sensor spatiotemporal calibration

When conducting experiments on intersection scenes, it is also necessary to calibrate the sensors in terms of time and space after arranging the sensors [25]. Time calibration is the inter frame time synchronization of heterogeneous sensors, which solves the problem of misalignment between radar frames and video frames caused by different sampling frequencies. Practical experiments commonly use linear interpolation method to achieve time alignment. In addition, time alignment can also be achieved by selecting the least common multiple of multi-sensor sampling times as the overall sampling time. When conducting experiments on the algorithm by a simulation platform in this paper, the sampling frequency can be normalized by setting the sensor_tick between the two types of sensors to the same value to achieve frame synchronization between sensors.

Space calibration refers to unifying the digitized coordinate systems used in the system, such as mmw radar, RGB cameras, image planes, etc. Schematic diagram of sensor coordinate system relationship is shown in Fig. 7.

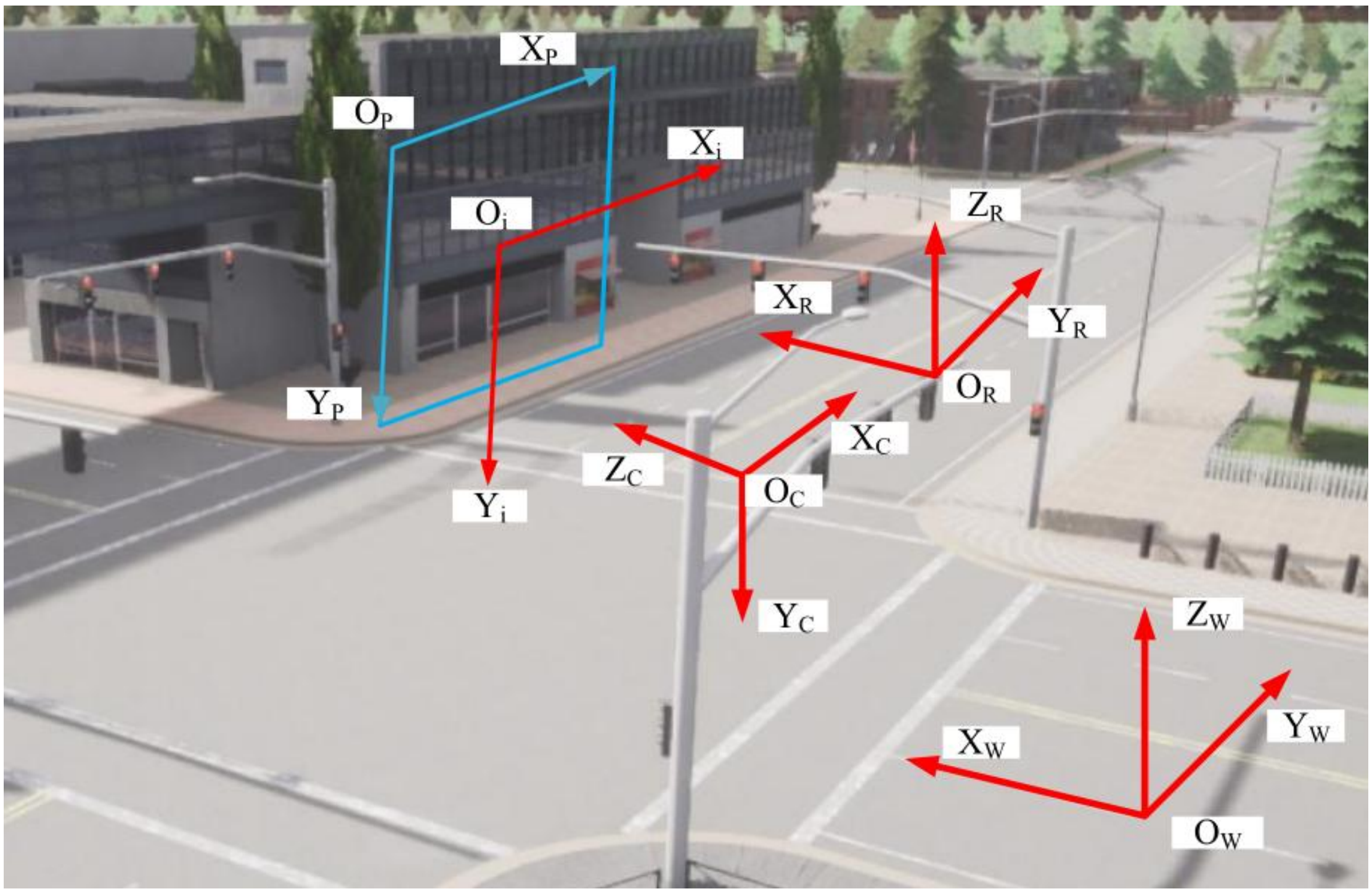

Fig. 7. Schematic diagram of sensor coordinate system relationship

where $O_W X_W Y_W Z_W$ represents world coordinate system. The origin and direction of the coordinate system are determined when creating a scene. $O_R X_R Y_R Z_R$ represents mmw radar coordinate system, $O_C X_C Y_C Z_C$ represents RGB cameras coordinate system, $O_i X_i Y_i$ represents image coordinate system obtained by cameras, and $O_P X_P Y_P$ represents pixel coordinate system of images. The spatial calibration of sensors is mainly to convert the target points collected by radar sensors into coordinate systems and project them onto the pixel plane, which is consistent with the images collected by visual sensors. Because of the fact that mmw radar generally outputs data in polar coordinates, including distance ρ, azimuth θ, and elevation φ. So, it is necessary to convert it to the Cartesian coordinate system first. The conversion formula is shown in (2), and then the coordinates of the target point P in the mmw radar coordinate system, $P_R = (x_R, y_R, z_R)^T$, can be obtained.

$$\begin{cases} x_R = \rho cos\theta cos\varphi \\ y_R = \rho sin\theta cos\varphi \\ z_R = \rho sin\varphi \end{cases} \tag{2}$$

The conversion between radar coordinate system and world coordinate system, as well as the conversion between camera coordinate system and world coordinate system, are both offset and rotation of Cartesian coordinate system. The offset and rotation parameters for converting radar coordinate system coordinates $P_R$ to world coordinate system coordinates $P_W = (x_W, y_W, z_W)^T$ can be obtained by installing and arranging millimeter wave radar at the position and direction. The offset and rotation parameters for converting coordinates from the camera coordinate system $P_C = (x_C, y_C, z_C)^T$ to the world coordinate system can be obtained through the camera extrinsic matrix. Taking the latter as an example, $P_C$ can be obtained by equation (3) and (4).

$$\begin{bmatrix} x_C \\ y_C \\ z_C \\ 1 \end{bmatrix} = \begin{bmatrix} R_{W,C} & T_{W,C} \\ 0 & 1 \end{bmatrix} \begin{bmatrix} x_W \\ y_W \\ z_W \\ 1 \end{bmatrix} \tag{3}$$

$$R_{W,C} = \begin{bmatrix} r_{11} & r_{12} & r_{13} \\ r_{21} & r_{22} & r_{23} \\ r_{31} & r_{32} & r_{33} \end{bmatrix}, T_{W,C} = \begin{bmatrix} t_x \\ t_y \\ t_z \end{bmatrix} \tag{4}$$

where $R_{W,C}$ represents the rotation matrix for converting the world coordinate system to the camera coordinate system, and it is a unit orthogonal matrix of 3×3. $T_{W,C}$ represents translation relationship between two coordinate systems.

As is shown in Fig. 8, the conversion from camera coordinate system to pixel coordinate system utilizes the pinhole imaging principle of the camera. Objects in three-dimensional space are projected onto the image plane $O_iX_iY_i$, where $f$ represents the focal length of the camera. The transformation relationship between $(x_i, y_i)$ and $(x_C, y_C, z_C)$ inferred from similar triangles is shown in (5). The mathematical expression for the transformation matrix from image coordinate system to pixel coordinate system is shown in (6).

$$\begin{cases} x_i = f\frac{x_C}{z_C} \\ y_i = f\frac{y_C}{z_C} \end{cases} \tag{5}$$

$$\begin{bmatrix} x_P \\ y_P \\ 1 \end{bmatrix} = \begin{bmatrix} \frac{1}{dx} & 0 & x_{P0} \\ 0 & \frac{1}{dy} & y_{P0} \\ 0 & 0 & 1 \end{bmatrix} \begin{bmatrix} x_i \\ y_i \\ 1 \end{bmatrix} \tag{6}$$

where $dx, dy$ represents the physical size of a pixel in x, y direction, $x_{P0}$ and $y_{P0}$ represents the coordinate of the camera's optical axis in the pixel plane, which is the main point coordinate of the camera.

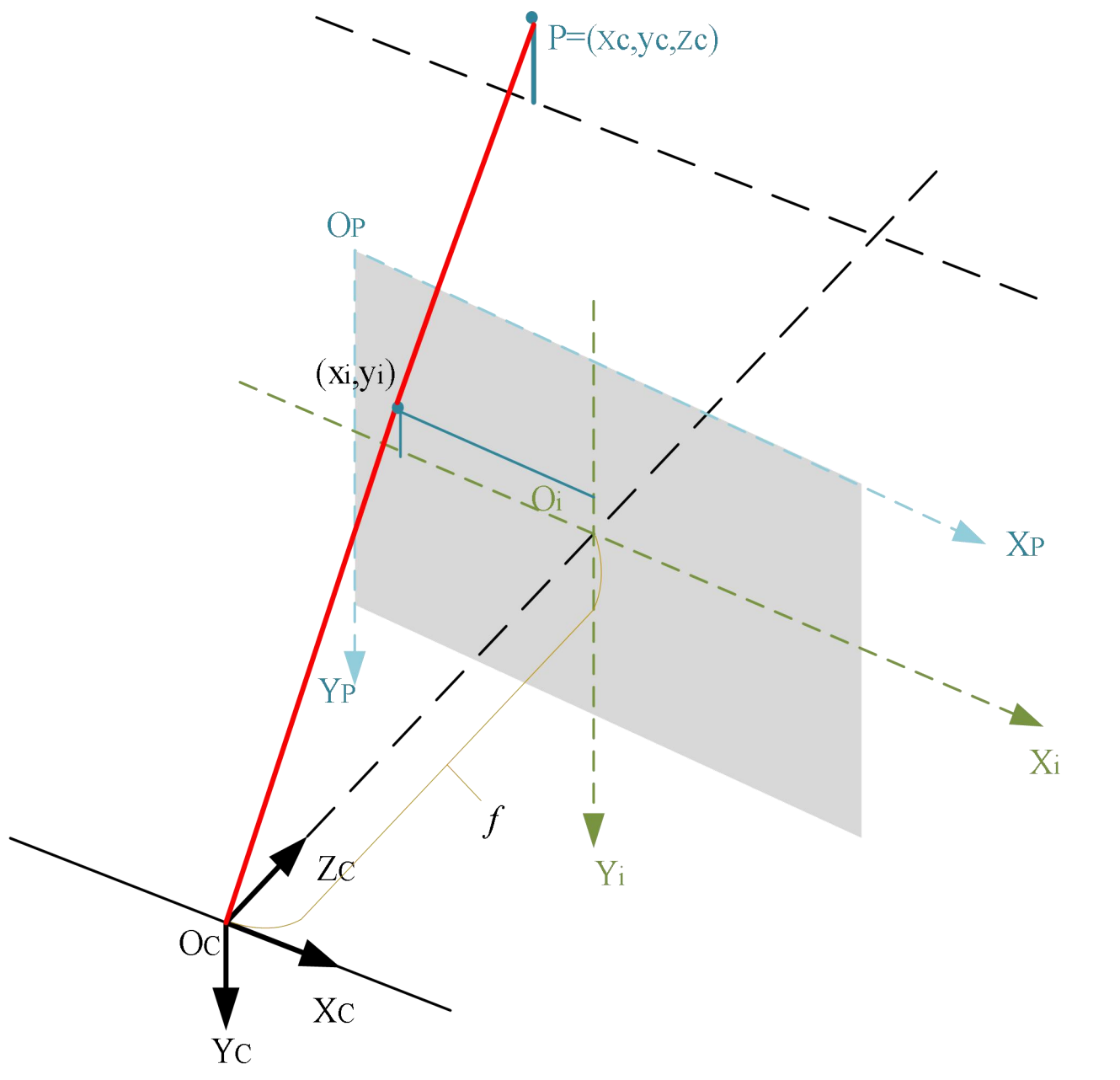

Fig. 8. Schematic diagram of camera imaging principle

According to the transformation relationship of the above coordinate system, the transformation relationship between the target point from mmw radar to pixel coordinate system is shown in (7):

$$z_C\begin{bmatrix}x_P\\y_P\\1\end{bmatrix}=\underbrace{\begin{bmatrix}\frac{1}{dx}&0&x_{P0}\\0&\frac{1}{dy}&y_{P0}\\0&0&1\end{bmatrix}\begin{bmatrix}f&0&0&0\\0&f&0&0\\0&0&1&0\end{bmatrix}}_{M_1}\underbrace{\begin{bmatrix}R_{W,C}&T_{W,C}\\0&1\end{bmatrix}}_{M_2}\underbrace{\begin{bmatrix}R_{R,W}&T_{R,W}\\0&1\end{bmatrix}}_{M_3}\begin{bmatrix}\rho\cos\theta\cos\varphi\\\rho\sin\theta\cos\varphi\\\rho\sin\varphi\\1\end{bmatrix} \tag{7}$$

where $R_{R,W}$ and $T_{R,W}$ represents rotation matrix and offset from conversion of mmw coordinate system to world coordinate system respectively, $M_1$ represents internal parameter matrix of camera sensors, $M_2$ represents camera external parameter matrix, and $M_3$ represents mmw radar conversion matrix.

B. RGB image modules of mmw radar

The multi-sensor fusion target detection algorithm inputs designed in this paper includes radar branches, so it is necessary to consider how to generate point cloud images of radar branch inputs. The outputs of mmw radar sensors includes four dimensions: velocity, distance, azimuth and elevation. Fig. 9(a) is a 3D scatter plot of point cloud data collected by a frame of radar. The arrangement position of mmw radar is at the origin of the scatter plot. The target detection algorithm proposed in this paper needs to convert it into a 2D image. After the coordinate transformation in section A, the point cloud can be projected onto the pixel plane, but this only includes the position information of the target, ignoring the physical state of the sensor output target, namely distance and velocity information. In this section, the radar image generation model is redesigned to convert the physical state of the target into pixel values according to certain quantization standards. The quantization conversion equation is shown in (8):

$$\begin{cases}R=\frac{128(d+20)}{250}+127\\G=\frac{128(v+40)}{50}+127\\B=\frac{128(v-20)}{50}+127\end{cases} \tag{8}$$

where $\mathrm{R,G,B}$ represents pixel values of three channels for converting radar point clouds to pixel planes respectively, d represents the distance from target point to radar, and v represents the radial velocity from target point to radar. Considering that the X axis direction of the arranged mmw radar always points straight ahead, which results in negative velocity values being collected. So, a quantization conversion standard shown in equation (8) was designed. The radar image in Fig. 9(b) shows the point cloud from Fig. 9(a) after coordinate transformation and adding physical information. The image size is set to 1080×1080, which is consistent with visual images. In areas without radar points, the pixel values of three channels are set to 0.

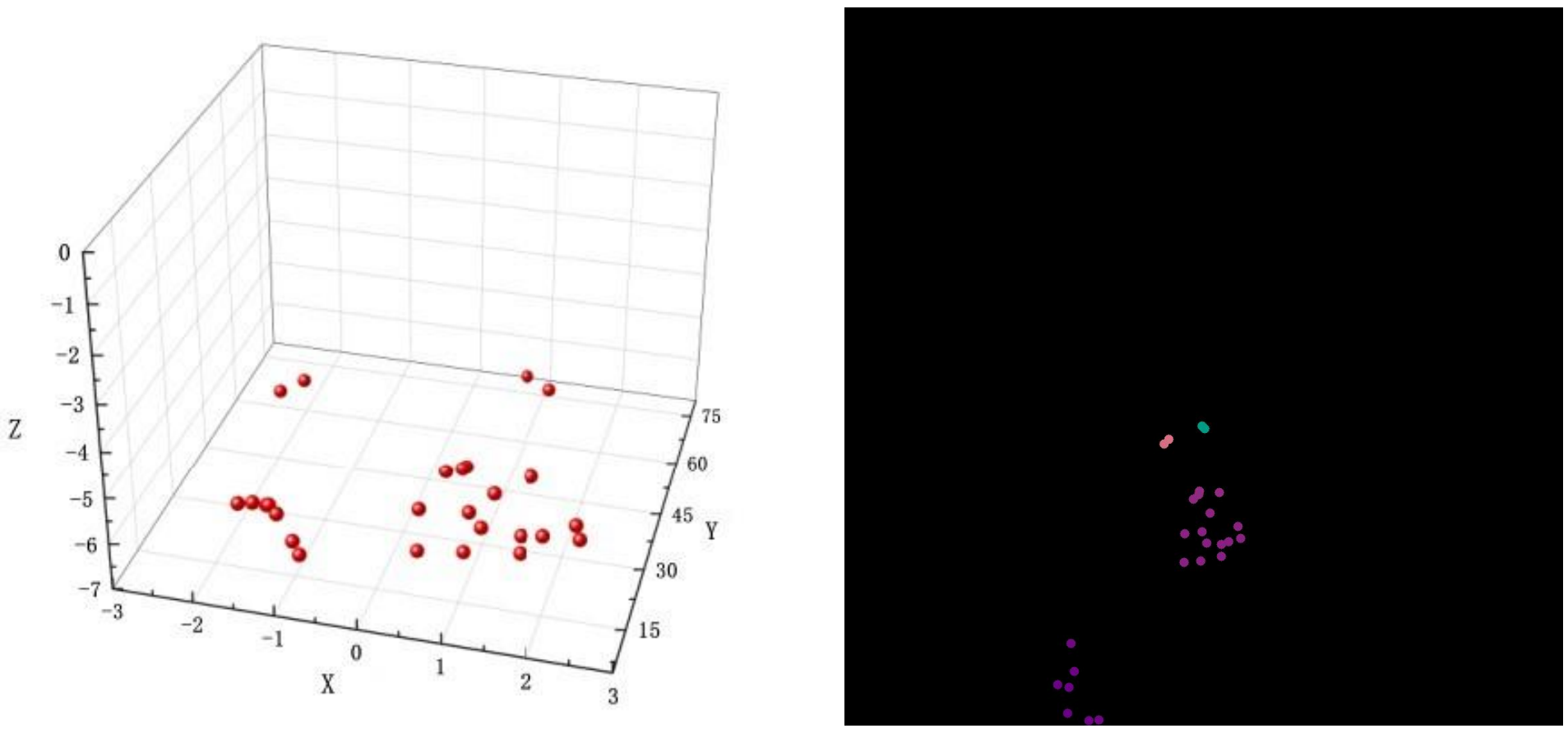

(a) (b)

Fig. 9. Millimeter-wave radar RGB image generation model: (a) Radar 3D scatter plot, (b) Radar RGB image

In the mmw radar RGB image model constructed in this paper, the position of the point and the position of the target in the real world also maintain the camera imaging relationship shown in Fig. 8, and its $R, G, B$ three channels also contain distance and velocity information of the target. Subsequent experiments all use this model to preprocess the radar images.

C. Making experimental dataset

The current mainstream traffic direction datasets, including KITTI [26], NuScenes [27], etc., are all applied in the research of autonomous driving algorithms. However, they are not applicable to the research of roadside sensors for traffic vehicle perception algorithms. So, this paper makes 2D dataset including original data of cameras and millimeter-wave radar.

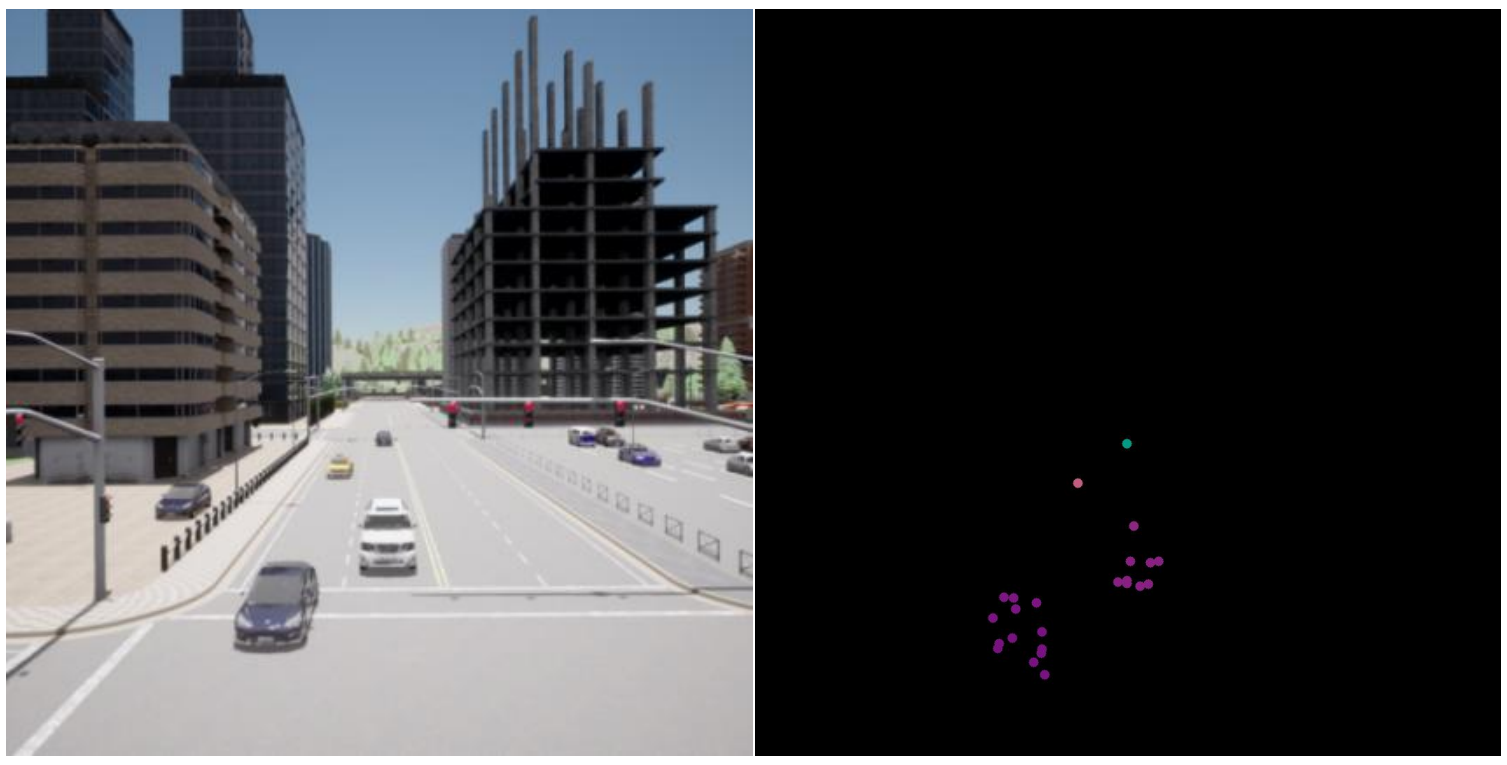

(a)

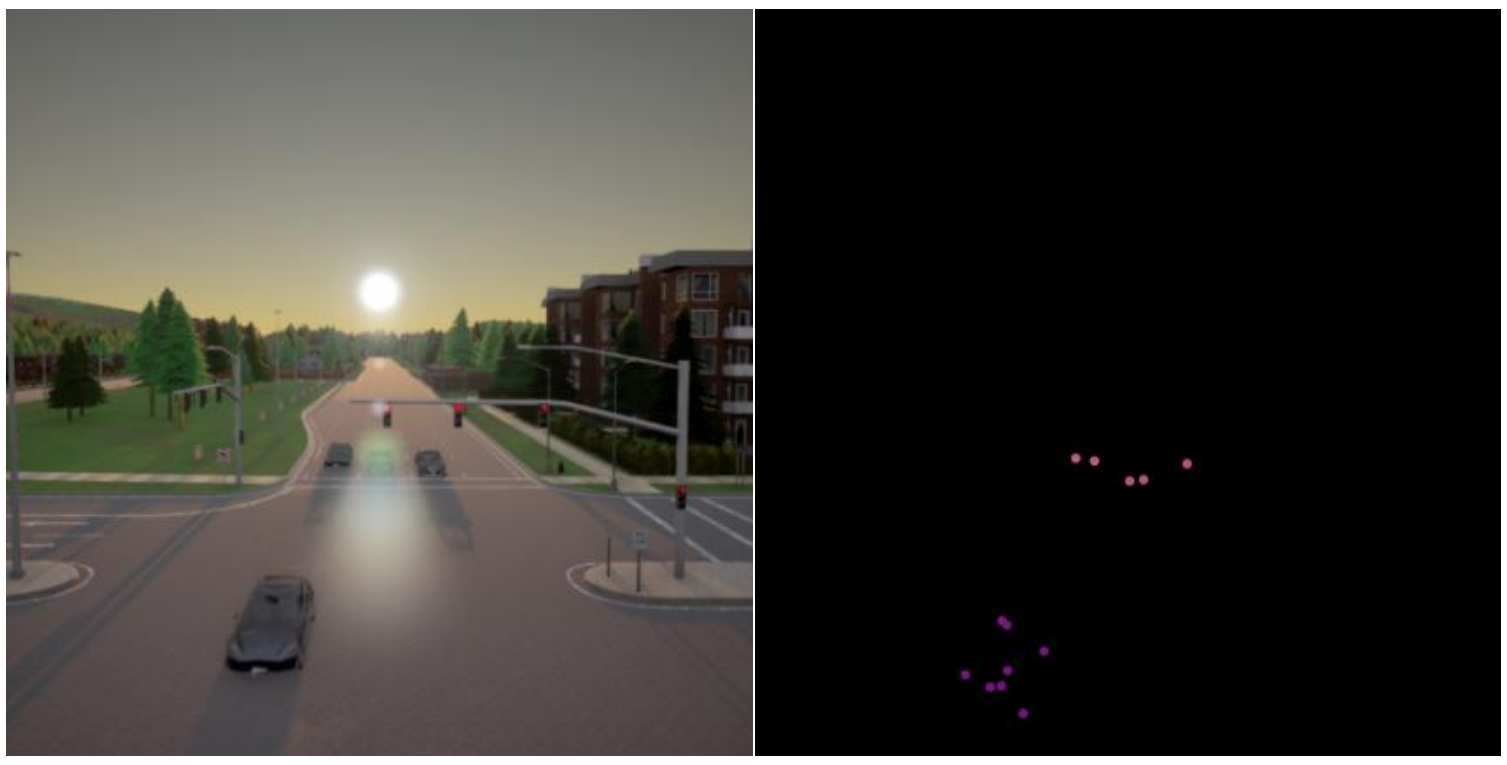

(b)

Fig. 10. Image example of the dataset: (a) Scene 1: visual image (left), mmw radar image (right), (b) Scene 2: visual image (left), mmw radar image (right).

The dataset used in this paper is obtained by customizing the scene, environment and traffic flow through autonomous driving simulation software. The dataset contains three sets of data: mmw radar point cloud images, visual images and label files. The first two sets of data are directly obtained by simulation software, and the label file is a JSON file manually marked and generated by labelme software. The dataset selected in this paper includes 16 traffic intersection scenes and a total of 3176 sets of images collected. Fig. 10 shows the images captured and collected by the camera and radar after adding traffic vehicles, which is an example of the image size in the dataset. All of the images are in size 1080×1080. This paper divides the dataset into training, validation and testing sets in a ratio of 5.5:2.5:2 randomly, with 1688 training sets, 848 validation sets, and 640 testing sets.

## V. EXPERIMENT AND ANALYSIS OF RV-PAFCOS TARGET DETECTION

In this section, the proposed RV-PAFCOS target detection model will be trained and tested on the basis of the dataset above. This experiment mainly consists of three parts. Firstly, try to replace the RV-PAFCOS lightning vision fusion module, including ADD, MUL, CAT, SAC, and conduct a large number of experiments under the same configuration parameters to obtain RV-PAFCOS of the best target detection effect. Then, compare the detection results between RV-PAFCOS of radar images and visual images and FCOS of only visual images. Finally, try different configurations of attention matrix convolution modules and compare their experimental results.

### A. Experimental environment and evaluation indicators

The software operating system used in this experiment is Ubuntu 18.04, and the specific software and hardware information is shown in Table II. In the experiment, the RV-PAFCOS network model was trained by Stochastic Gradient Descent (SGD) [28], with momentum set to 0.9, weight decay set to 0.0001, and learning rate initialized to 0.001. Use the MSRA [29] method to initialize the weights. A total of 40000 iterations were conducted, with 4 pairs of radar and visual images as inputs for each iteration, totaling 8 images. During the training process, the size of the input images was adjusted. In this experiment, the image size was adjusted to 800×800.

TABLE II

NETWORK TRAINING EXPERIMENT ENVIRONMENT

| Operating system | Ubuntu | 18.04 |
|---|---|---|
| Software configuration | Python | 3.7.0 |
| | Pytorch | 1.10.1 |
| | CUDA | 10.2 |
| Hardware configuration | CPU | Intel(R) Xeon(R) E5-2620 v3 |
| | Graphics card | NVIDIA GeForce RTX 3060 |
| | Memory | 64GB |

The dataset format in this experiment is the same as the MS COCO dataset, including training set, validation set, testing set, and annotation files of JSON format. The annotation files consist of fields such as image number, category number, and object bounds coordinates. In addition, the quantitative analysis indicators used are consistent with the COCO dataset, mainly including Average Precision (AP) and Average Recall (AR). There are six detailed classifications of AP, including $AP^{IoU=0.5}$, $AP^{IoU=0.75}$, AP, $AP^{small}$, $AP^{medium}$ and $AP^{large}$. The first two are the average precision when $IoU$ is equal to 0.5 and 0.75 respectively. The third one refers to the average value of all APs with IoU in the range of 0.5 to 0.95, with a step of 0.05. The last three are to evaluate the detection performance of different target scales. Among all detection indicators, the maximum number of detection boxes for each image is set to 100, so the subsequent abbreviations are $AP^{.50}(100)$, $AP^{s}(100)$, etc. Under different detection box numbers and scales, AR indicators are classified into $AR(1)$, $AR(10)$, $AR(100)$, $AR^{s}(100)$, $AR^{m}(100)$ and $AR^{l}(100)$ in detail.

B. Experimental comparisons of different fusion modules

In this section, the four fusion methods proposed in Section 3B are first evaluated, and the experimental environment and parameters are consistent with those described in Section 5A. Because of the need for the CAT and SAC modules to cascade the outputs of two image branches by channel, the number of channels increases. To ensure that the four methods have the same number of channels after going through the fusion module, an additional Conv 512, 256, 1×1, 1 needs to be added, which means a convolutional layer with 1×1 convolution kernel size and a step size of 1. This can ensure that the number of output channels for all four methods is 256.

Table III shows the average precision and average recall of four fusion modules detected on RV-PAFCOS. On each indicator, the optimal detection results are displayed in bold. Because of the fact that there were no more than ten effective targets in each image of the dataset used in this experiment, the results of $AR(10)$ and $AR(100)$ were always consistent. From the table, it can be seen that RV-PAFCOS using the SAC fusion module is superior to other fusion modules in almost all AP and AR indicators. In the AP $(100)$ indicator, it is about 9.1%, 6.6% and 12.9% higher than MUL, ADD and CAT respectively. Similar to references [23], [30], the additive fusion scheme is superior to the cascaded fusion scheme, and even performs the best on $AP^{1}(100)$, but other performance is slightly lower than SAC. Currently, the SAC module that uses convolutional blocks 1×1, 3×3 and 5×5 to extract attention weight matrices is the optimal solution for target detection performance.

TABLE III

AP AND AR OF DIFFERENT FUSION MODULES DETECTED ON RV-PAFCOS

| Model | Fusion module | $AP(100)$ | $AP^{.50}(100)$ | $AP^{.75}(100)$ | $AP^{s}(100)$ | $AP^{m}(100)$ | $AP^{1}(100)$ |
|---|---|---|---|---|---|---|---|
| RV-PAFCOS | MUL | 63.7 | 96.1 | 76.1 | 49.9 | 70.9 | 71.9 |
| | ADD | 66.2 | 97.3 | 79.1 | 51.9 | 71.3 | **79.4** |
| | CAT | 59.9 | 95.8 | 68.2 | 44.6 | 67.5 | 72.6 |
| | SAC | **72.8** | **98.8** | **89.7** | **63.5** | **78.1** | 77.1 |
| Model | Fusion module | $AR(1)$ | $AR(10)$ | $AR(100)$ | $AR^{s}(100)$ | $AR^{m}(100)$ | $AR^{1}(100)$ |
| RV-PAFCOS | MUL | 26.7 | 68.6 | 68.6 | 58.2 | 75.4 | 76.1 |
| | ADD | 27.6 | 70.9 | 70.9 | 61.1 | 75.9 | **82.4** |
| | CAT | 26.1 | 65.7 | 65.7 | 53.4 | 72.6 | 78.1 |
| | SAC | **29.1** | **77.4** | **77.4** | **70.7** | **82.1** | 80.8 |

C. Comparison between RV-PAFCOS and FCOS

This section uses a fusion module to compare RV-PAFCOS of SAC with FCOS trained only on visual images. The fusion module consists of three convolutional layers, which are used to extract spatial attention matrices. Attention weights can control or enhance the information of visual images. Fig. 11 shows the Loss curve and AP curve of two networks after 40000 iterations of training, with the horizontal axis representing the number of iterations. The data for the AP curve is obtained through separate testing every 2500 iterations. From Fig. 11, it can be seen that throughout the entire iteration process, the training loss of RV-PAFCOS decreases faster than FCOS and the final convergence is also lower. In addition, it can be seen from the AP curve that the AP accuracy of RV-PAFCOS quickly stabilizes and remains higher than FCOS throughout the entire iteration process.

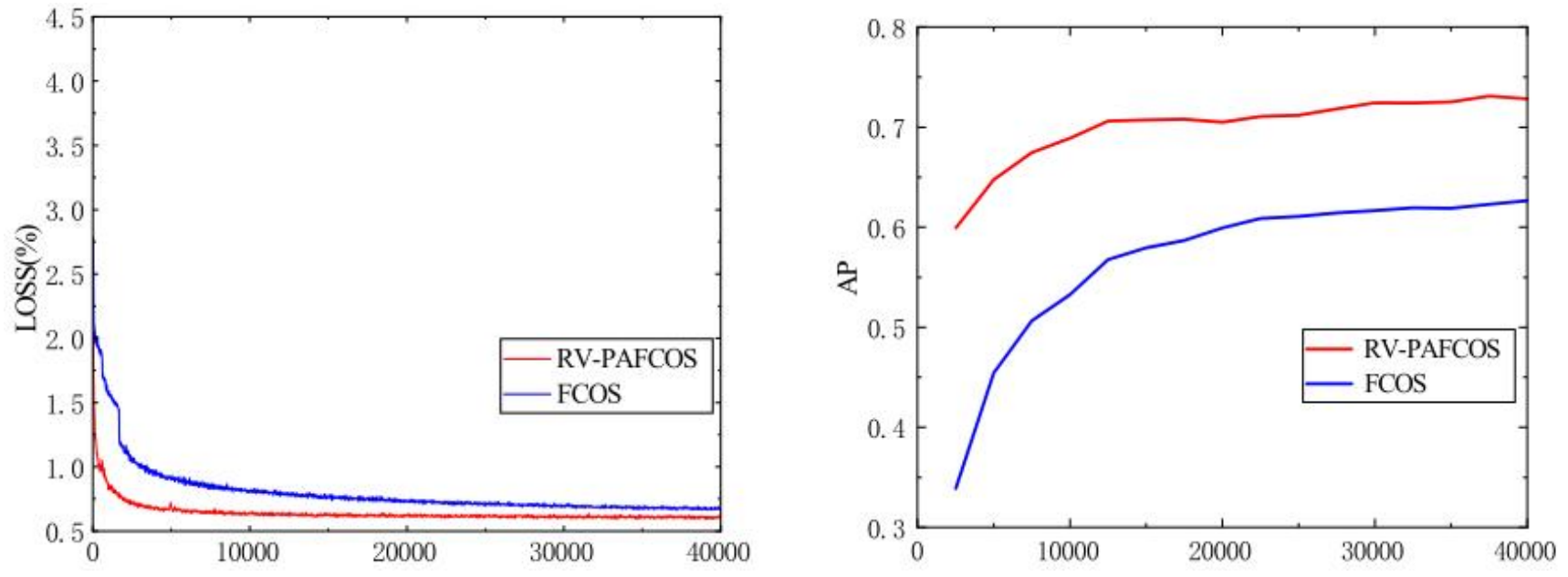

(a) (b)

Fig. 11. Comparison of RV-PAFCOS and FCOS training process: (a) Loss curve, (b) AP curve.

For two detection networks, quantitative analysis was conducted by two types of indicators: average precision and average recall. The experimental results are shown in Table IV, indicating that the average precision of RV-PAFCOS is relatively better than FCOS under all indicators. In AP(100), AP.50(100) and AP.75(100), RV-PAFCOS has advantages of 10.2%, 1.3%, and 15.7% respectively, indicating that the bounding boxes generated by RV-PAFCOS for target detection are more compact and accurate. After adding radar images, the network helps with the identification of small and medium-sized targets, and improves the average precision of detection for small and medium-sized targets. In addition, the average recall rate of RV-PAFCOS is better than FCOS under all scales and indicators.

TABLE IV

AP AND AR OF TRAINED RV-PAFCOS AND FCOS

| Model | Fusion module | AP(100) | $AP^{.50}(100)$ | $AP^{.75}(100)$ | $AP^{s}(100)$ | $AP^{m}(100)$ | $AP^{l}(100)$ |
|---|---|---|---|---|---|---|---|
| FCOS | None | 62.6 | 97.5 | 74.0 | 53.2 | 68.7 | 64.6 |
| RV-PAFCOS | SAC | **72.8** | **98.8** | **89.7** | **63.5** | **78.1** | **77.1** |
| Model | Fusion module | AR(1) | AR(10) | AR(100) | $AR^{s}(100)$ | $AR^{m}(100)$ | $AR^{l}(100)$ |
| FCOS | None | 26.3 | 69.4 | 69.4 | 62.4 | 75.4 | 70.0 |
| RV-PAFCOS | SAC | **29.1** | **77.4** | **77.4** | **70.7** | **82.1** | **80.8** |

Fig. 12 shows the qualitative results visualized in this comparison, with the left column showing the FCOS detection results and the right column showing the detection results of RV-PAFCOS. Compared with FCOS, RV-PAFCOS can detect traffic vehicles that cannot be detected by visual sensors because of environmental and climate influences. The addition of information collected by mmw radar during the detection process breaks through the limitation of detection algorithms only relying on visual sensors, and improves the accuracy and robustness of a single visual sensor detection system.

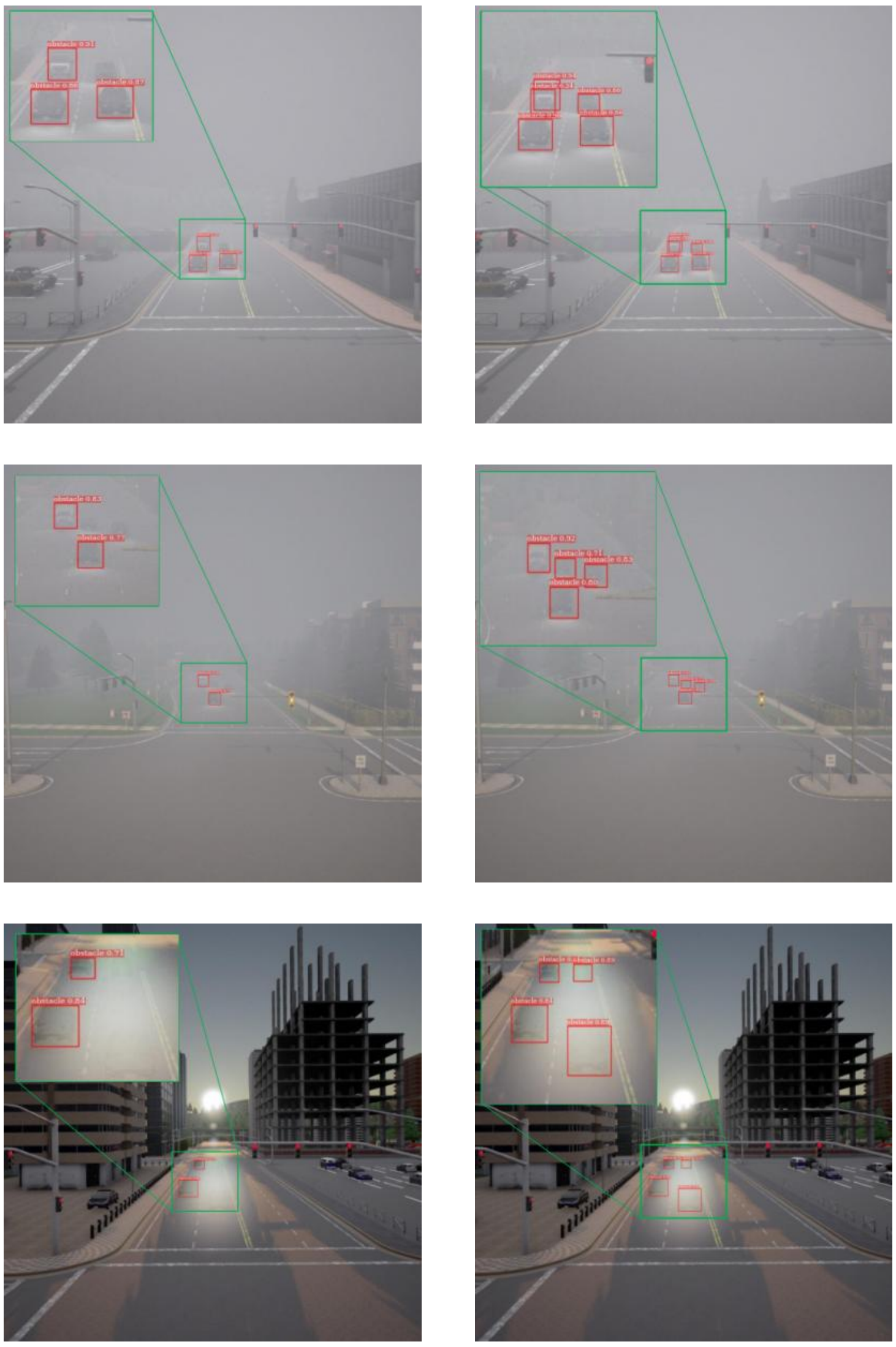

Fig. 12. FCOS (left) and RV-PAFCOS (right) detection visualization results

D. Experimental comparisons under different configurations of SAC

According to the experimental analysis in sections B and C, it can be concluded that the detection performance of RV-PAFCOS using the SAC fusion module is superior to other fusion modules. This section will modify the configuration of the SAC module by adjusting the convolutional layer configuration for extracting attention matrices and try different convolutional kernel sizes and combinations. Finally, this section will conduct experiments in an environment where other network parameters and training hyper-parameters are consistent and analyze its detection performance.

The sizes of the convolution kernels used in the experiment are 1×1, 3×3, 5×5, 7×7, 9×9. Seven experiments were conducted using different arrangement and combination methods. Table V compares the average precision and average recall of RV-PAFCOS network under SAC fusion modules with different convolutional kernel configurations.

TABLE V

DETECTION PERFORMANCE OF RV-PAFCOS USING SAC MODULES WITH DIFFERENT CONFIGURATIONS

| SAC module configuration for RV-PAFCOS 1×1 | 3×3 | 5×5 | 7×7 | 9×9 | AP(100) | $AP^{.50}$(100) | $AP^{.75}$(100) | $AP^{s}$(100) | $AP^{m}$(100) | $AP^{l}$(100) |
|---|---|---|---|---|---|---|---|---|---|---|
| √ | √ | | | | 70.6 | 98.6 | 88.2 | 62.2 | 76.4 | 76.6 |
| | √ | √ | | | 70.8 | 98.7 | 88.5 | 60.2 | 76.1 | 77.2 |
| | | √ | √ | | 70.9 | 98.6 | 86.5 | 60.8 | 77.2 | 75.3 |
| | | | √ | √ | 71.1 | 98.7 | 88.8 | 63.0 | 76.2 | 77.0 |
| √ | √ | √ | | | **72.8** | **98.8** | **89.7** | **63.5** | **78.1** | 77.1 |
| | √ | √ | √ | | 71.8 | 98.7 | 88.4 | 60.8 | 77.3 | **79.8** |
| | | √ | √ | √ | 72.6 | 98.7 | **89.7** | 62.6 | 78.0 | 77.5 |
| **SAC module configuration for RV-PAFCOS** 1×1 | 3×3 | 5×5 | 7×7 | 9×9 | AR(1) | AR(10) | AR(100) | $AR^{s}$(100) | $AR^{m}$(100) | $AR^{l}$(100) |
| √ | √ | | | | 28.6 | 75.2 | 75.2 | 68.8 | 80.5 | 80.4 |
| | √ | √ | | | 28.4 | 75.8 | 75.8 | 68.1 | 81.0 | 80.3 |
| | | √ | √ | | 28.6 | 76.1 | 76.1 | 67.9 | 82.2 | 79.4 |
| | | | √ | √ | 28.6 | 76.8 | 76.8 | 70.5 | 81.2 | 80.3 |
| √ | √ | √ | | | **29.1** | **77.4** | **77.4** | **70.7** | **82.1** | **80.8** |
| | √ | √ | √ | | 28.6 | 77.0 | 77.0 | 69.1 | 81.9 | 80.5 |
| | | √ | √ | √ | 29.0 | **77.4** | **77.4** | 70.3 | 81.7 | 81.5 |

For the SAC fusion module composed of two convolutional layers, the performance of the detection network is very similar, with an average precision of around 70.8%. For the SAC fusion module composed of three convolutional layers, the detection performance is better than using two convolutional layers. The three convolutional layers can provide more scale feature information, and the generated spatial attention matrix has a greater positive effect on the fusion of the two sensors. From the table, it can be seen that the training network using the SAC fusion module composed of 1×1, 3×3 and 5×5 convolutional layers is almost superior or equivalent to other schemes in all indicators.

## VI. CONCLUSION

This paper proposed a traffic intersection vehicle detection method RV-PAFCOS based on the fusion of mmw radar and visual sensors, which performs inter class fusion of multi-sensors from the term of feature fusion. Compared with other fusion modules, the SAC fusion method proposed in this paper fully utilizes radar features under different levels and scales. The generated spatial attention weight matrix can control or enhance the image information of visual sensors more effectively. Through experiments, it has been verified that the detection performance of RV-PAFCOS using the SAC fusion module is better under all scale evaluation systems. In addition, the impact of different configurations of SAC modules on the target detection network was compared through experiments, and RV-PAFCOS with the best detection performance was selected. This detection method will use CARLA as the development platform to achieve a traffic intersection vehicle detection system based on Radar-Vision fusion.

## REFERENCES

[1] D. G. Costa. Visual Sensors Hardware Platforms: A Review [J]. IEEE Sensors Journal, 2020, 20 (8): 4025-4033.

[2] F. Cui et al. Online Multipedestrian Tracking Based on Fused Detections of Millimeter Wave Radar and Vision [J]. IEEE Sensors Journal, 2023, 23 (14): 15702-15712.

[3] M. Zong, Z. Zhu and H. Wang. A Simulation Method for Mmw Radar Sensing in Traffic Intersection Based on Bidirectional Analytical Ray-Tracing Algorithm [J]. IEEE Sensors Journal, 2023, 23 (13): 14276-14284.

[4] Russell M E, Crain A, Curran A, et al. Millimeter-wave radar sensor for automotive intelligent cruise control (ICC) [J]. IEEE Transactions on Microwave Theory and Techniques, 1997, 45 (12): 2444-2453.

[5] Lei C, Zhao C, Ji Y, et al. Identifying and correcting the errors of vehicle trajectories from roadside millimetre-wave radars [J]. IET Intelligent Transport Systems, 2023, 17 (2): 418-434.

[6] Z. Wang, X. Xie, Q. Zhao and G. Shi. Filter Clustering for Compressing CNN Model With Better Feature Diversity [J]. IEEE Transactions on Circuits and Systems for Video Technology, 2023, 33 (12): 7385-7397.

[7] Hsu Y-W, Lai Y-H, Zhong K-Q, et al. Developing an on-road object detection system using monovision and radar fusion [J]. Energies, 2019, 13 (1): 116.

[8] Ravindran R, Santora M J, Jamali M M. Multi-object detection and tracking, based on DNN, for autonomous vehicles: A review [J]. IEEE Sensors Journal, 2020, 21 (5):5668-5677.

[9] Palffy A, Dong J, Kooij J F, et al. CNN based road user detection using the 3D radar cube [J]. IEEE Robotics and Automation Letters, 2020, 5 (2): 1263-1270.

[10] M. Alameh, Y. Abbass, A. Ibrahim, G. Moser and M. Valle. Touch Modality Classification Using Recurrent Neural Networks [J]. IEEE Sensors Journal, 2021, 21(8): 9983-9993.

[11] F. Cui et al. Online Multipedestrian Tracking Based on Fused Detections of Millimeter Wave Radar and Vision [J]. IEEE Sensors Journal, 2023, 23(14): 15702-15712.

[12] A. Gupta and A. Choudhary. A Framework for Camera-Based Real-Time Lane and Road Surface Marking Detection and Recognition [J]. IEEE Transactions on Intelligent Vehicles, 2018, 3(4): 476-485.

[13] He K, Zhang X, Ren S, et al. Spatial pyramid pooling in deep convolutional networks for visual recognition [J]. IEEE Transactions on Pattern Analysis and Machine Intelligence,2015, 37 (9): 1904-1916.

[14] Girshick R. Fast r-cnn [C]. IEEE International Conference on Computer Vision, 2015:1440-1448.

[15] Ren S, He K, Girshick R, et al. Faster r-cnn: Towards real-time object detection with region proposal networks [J]. Advances in Neural Information Processing Systems,2015, 28.

[16] Dai J, Li Y, He K, et al. R-fcn: Object detection via region-based fully convolutional networks [J]. Advances in Neural Information Processing Systems, 2016, 29.

[17] Y. Song, Z. Xie, X. Wang and Y. Zou. MS-YOLO: Object Detection Based on YOLOv5 Optimized Fusion Mmw Radar and Machine Vision [J]. IEEE Sensors Journal, 2022, 22(15):15435-15447.

[18] X. Liang, J. Zhang, L. Zhuo, Y. Li and Q. Tian. Small Object Detection in Unmanned Aerial Vehicle Images Using Feature Fusion and Scaling-Based Single Shot Detector With Spatial

Context Analysis [J]. IEEE Transactions on Circuits and Systems for Video Technology, 2020, 30(6): 1758-1770.
[19] Tian Z, Shen C, Chen H, et al. Fcos: A simple and strong anchor-free object detector [J].IEEE Transactions on Pattern Analysis and Machine Intelligence, 2020, 44 (4): 1922-1933.
[20] N. Liu, J. Chen, H. Wu, F. Li and J. Gao. Microseismic First-Arrival Picking Using Fine-Tuning Feature Pyramid Networks [J]. IEEE Geoscience and Remote Sensing Letters, 2022,19:1-5.
[21] Y. Wu, Q. Yao, X. Fan, M. Gong, W. Ma and Q. Miao. PANet: A Point-Attention Based Multi-Scale Feature Fusion Network for Point Cloud Registration [J]. IEEE Transactions on Instrumentation and Measurement, 2023, 72:1-13.
[22] U. Saeed et al. Portable UWB RADAR Sensing System for Transforming Subtle Chest Movement Into Actionable Micro-Doppler Signatures to Extract Respiratory Rate Exploiting ResNet Algorithm [J]. IEEE Sensors Journal, 2021, 21(20):23518-23526.
[23] Chadwick S, Maddern W, Newman P. Distant vehicle detection using radar and vision [C]. International Conference on Robotics and Automation (ICRA), 2019: 8311-8317.
[24] John V, Mita S. RVNet: Deep sensor fusion of monocular camera and radar for image-based obstacle detection in challenging environments [C]. Image and Video Technology: 9th Pacific-Rim Symposium, PSIVT 2019, Sydney, NSW, Australia, November 18-22, 2019, Proceedings 9, 2019: 351-364.
[25] E. Wise, Q. Cheng and J. Kelly. Spatiotemporal Calibration of 3-D Millimetre-Wavelength Radar-Camera Pairs [J]. IEEE Transactions on Robotics, 2023, 39(6): 4552-4566.
[26] Geiger A, Lenz P, Stiller C, et al. Vision meets robotics: The kitti dataset [J]. The International Journal of Robotics Research, 2013, 32(11): 1231-1237.
[27] W. K. Fong et al. Panoptic Nuscenes: A Large-Scale Benchmark for LiDAR Panoptic Segmentation and Tracking [J]. IEEE Robotics and Automation Letters, 2022, 7(2): 3795-3802.
[28] Y. Lei, T. Hu, G. Li and K. Tang. Stochastic Gradient Descent for Nonconvex Learning Without Bounded Gradient Assumptions [J]. IEEE Transactions on Neural Networks and Learning Systems, 2020, 31(10):4394-4400.
[29] A. T. Abebe and C. G. Kang. Multi-Sequence Spreading Random Access (MSRA) for Compressive Sensing-Based Grant-Free Communication [J]. IEEE Transactions on Communications, 2021, 69(11):7531-7543.
[30] Chang S, Zhang Y, Zhang F, et al. Spatial attention fusion for obstacle detection using mmwave radar and vision sensor [J]. Sensors, 2020, 20 (4): 956.